# Cache Optimization for Memory Intensive Workloads on Multi-socket Multi-core servers

**Murthy Durbhakula**

**Indian Institute of Technology Hyderabad, India**

cs15resch11013@iith.ac.in, murthy.durbhakula@gmail.com

**Abstract.** Major chip manufacturers have all introduced multicore microprocessors. Multi-socket systems built from these processors are used for running various server applications. Depending on the application that is run on the system, remote memory accesses can impact overall performance. This paper presents a cache optimization that can cut down remote DRAM accesses. By keeping track of remote cache lines loaded from remote DRAM and by biasing the cache replacement policy towards such remote DRAM cache lines the number of cache misses are reduced. This in turn results in improvement of overall performance. I present the design details in this paper. I do a qualitative comparison of various solutions to the problem of performance impact of remote DRAM accesses. This work can be extended by doing a quantitative evaluation and by further refining cache optimization.

## 1 Introduction

Many commercial server applications today run on cache coherent NUMA (ccNUMA) based multi-socket multi-core servers. Depending on the application, DRAM accesses can impact overall performance; particularly remote DRAM accesses. These are inherent to the application. One way to ameliorate this problem is to rewrite the application. Another way is to bias the cache replacement policy towards remote DRAM accesses. In this paper, we present such a cache replacement policy which will adaptively bias its replacement towards remote DRAM lines, when it's helpful for performance. We present a mechanism to observe the usefulness of bias and use it only when it's helpful. For instance, if there is a phase of application where there is little spatial/temporal locality of remote cache lines then we turn off the bias. The rest of the paper is organized as follows: Section 2 presents the cache replacement policy. Section 3 presents adaptive mechanism to turn on and turn off the cache replacement policy bias. Section 4 briefly describes the qualitative methodology I used in evaluation. Section 5 presents results. Section 6 describes related work and Section 7 presents conclusions.

## 2 Cache replacement policy

In a ccNUMA system for every cache line there is a notion of home node. Home node is defined as the node in whose DRAM a dirty cache line gets written back when it gets replaced. There are various ways home node gets determined for a given cache line. For instance, the upper bits of physical address of cache line can determine the home node. Or it can be determined using a first-touch policy. In this paper we assume that home node is determined by uppers bits of cache-line physical address.

In any level of cache in a given node/socket we maintain a "remote-line-counter" per cache set which indicates the number of times a remote line did not get replaced due to bias. When we are selecting a line for replacement we can see from the upper bits of the tag whether the home node of the line selected for replacement is same as the node where the cache is. If they both don't match then it's a remote line. We then see if the remote-line-counter is greater than a set threshold H, for instance half the cache associativity. If it is not then we increment the counter and skip selecting this line for replacement and replace a local line. If the counter is greater than the threshold then we replace the remote cache line and set the counter back to zero. If all lines in a given set are remote lines then we replace a remote line which is in LRU position.

## 3 Adaptive mechanism to turn on/off the cache replacement bias

For every core in the system we maintain a metric called "Remote_Miss_Fraction".

In a given time window T we calculate Remote_Miss_Fraction = (Number of Remote misses)/(Number of total misses). We use low and high water-mark for this metric. If this metric exceeds a high water mark then we turn on the cache bias. High water mark can be, for instance, 0.5 and low water mark can be 0.1. We continue to enable bias as long as Remote_Miss_Fraction does not go below low water-mark. If it goes below low water-mark then we turn off the bias.

## 4 Methodology

I am using a qualitative methodology to compare the contributions of this paper with other existing approaches. Particularly I compared with hardware solutions and software solutions and the metrics I used are:

i) Need software support: That is, does the approach require changes from software or will it work seamlessly with existing software.

ii) Flexibility: That is, can the idea be improved or configured later on. Either software or hardware/software hybrid solutions have this advantage

iii) Verification complexity: Though hardware solutions work seamlessly with existing software they generally have verification complexity. They need to be fully verified before they can ship. Whereas software solutions can be potentially patched.

This work can be extended by doing a quantitative evaluation with various workloads.

## 5 Results

### 5.1 Hardware solutions:

**Remote cache in local DRAM**

Stanford FLASH project proposed using a portion of local DRAM as a cache for remote DRAM accesses, known as remote-access-cache (RAC) [1]. Similar to our solution this is a hardware solution that does not require any changes in software. However, if the remote cache line working set is small then our approach of biasing cache replacement towards remote lines will let the remote line live in cache hence resulting in lesser fetch latency than fetching the same remote line from DRAM. Hence our solution works better for small to medium working set, since we bias cache replacement at every cache level, and local DRAM based remote cache works for really large working sets.

## 5.2 Software solutions:

### Page replication and migration

OS based page replication and migration [2] has been proposed as a software solution for this problem. For a read-only page such as code pages we can replicate the page at multiple nodes thus converting them from remote DRAM accesses to local DRAM accesses. Further if a data page is going to be written by threads of only one node then that page can be migrated to the node where threads reside. However, if the same page is going to be written by multiple threads of different nodes around same time then we cannot migrate the page. Similar to remote DRAM cache this solution works better for larger remote line working sets. For medium to small working sets biasing cache replacement results in lower access latency.

### OS Scheduling optimization

In my earlier work [3] I proposed optimizing OS scheduling algorithms to schedule such a way that the number of remote DRAM accesses are minimized for a thread. Again this will work for large working sets. When the remote line working set is medium to small, biasing cache replacement policy provides better access latency to remote lines.

| Solution | Need software changes | Flexibility | Hardware Verification complexity |
|---|---|---|---|
| Cache replacement policy based solution | No | No | Yes |
| Remote Access Cache | No | No | Yes |
| OS based page migration and replication | Yes | Yes | No |
| OS scheduling optimization | Yes | Yes | No |

**Table 1: Comparison of Various Solutions**

## 6 Related work

I have already discussed some related work in the previous section. In this section I am going to discuss some more. Srikanthan et al [4] proposed changes to the operating system to reduce both remote communication and remote DRAM misses. Whereas our solution is a pure hardware solution which works seamlessly with existing software. In cache-only-memory-architecture (COMA) [5] all of local DRAM is treated as a cache. Like ours this is a hardware solution. However for small to medium working sets our approach works better as we will be able to serve remote data directly from cache instead of local DRAM.

## 7 Conclusions

Many commercial server applications today run on cache coherent NUMA (ccNUMA) based multi-socket multi-core servers. Depending on the application, DRAM accesses can impact overall performance; particularly remote DRAM accesses. These are inherent to the application. In this paper we have presented a new cache replacement policy based solution that can help reduce the performance impact of remote DRAM accesses. We also presented an adaptive mechanism to observe the usefulness of bias and use it only when it's helpful. We have presented a qualitative evaluation of our approach. The solution presented in this paper does not require any software modifications. It works seamlessly with existing software.